\documentclass[aps,twocolumn,amsmath,amssymb,showpacs,prl,superscriptaddress,unsortedaddress]{revtex4}

\usepackage{epsf}
\usepackage{graphicx}

\newcommand{\etal}{{\it et al.}}

\begin{document}

\title{Momentum dependence of the superconducting gap in NdFeAsO$_{0.9}$F$_{0.1}$ single crystals measured by angle resolved photoemission spectroscopy}

\author{Takeshi~Kondo}
\affiliation{Ames Laboratory and Department of Physics and
Astronomy, Iowa State University, Ames, Iowa 50011, USA}

\author{A.~F.~Santander-Syro}
\affiliation{Laboratoire Photons Et Mati\`{e}re, UPR-5 CNRS, ESPCI, 10 rue Vauquelin, 75231 Paris cedex 5, France}
\affiliation{Labratoire de Physique des Solides, UMR-8502 CNRS, Universit\'{e} Paris-Sud, B\^{a}t. 510, 91405 Orsay, France}

\author{O.~Copie}
\affiliation{Unit\'e Mixte de Physique CNRS/Thales, Route d\'epartementale 128, 91767 Palaiseau Cedex, France}

\author{Chang~Liu}
\affiliation{Ames Laboratory and Department of Physics and
Astronomy, Iowa State University, Ames, Iowa 50011, USA}

\author{M.~E.~Tillman}
\affiliation{Ames Laboratory and Department of Physics and
Astronomy, Iowa State University, Ames, Iowa 50011, USA}

\author{E.~D.~Mun}
\affiliation{Ames Laboratory and Department of Physics and
Astronomy, Iowa State University, Ames, Iowa 50011, USA}

\author{J.~Schmalian}
\affiliation{Ames Laboratory and Department of Physics and
Astronomy, Iowa State University, Ames, Iowa 50011, USA}

\author{S.~L.~Bud'ko}
\affiliation{Ames Laboratory and Department of Physics and
Astronomy, Iowa State University, Ames, Iowa 50011, USA}

\author{M.~A.~Tanatar}
\affiliation{Ames Laboratory and Department of Physics and
Astronomy, Iowa State University, Ames, Iowa 50011, USA}

\author{P.~C.~Canfield}
\affiliation{Ames Laboratory and Department of Physics and
Astronomy, Iowa State University, Ames, Iowa 50011, USA}

\author{A.~Kaminski}
\affiliation{Ames Laboratory and Department of Physics and
Astronomy, Iowa State University, Ames, Iowa 50011, USA}

\date{\today}

\begin{abstract}
We use angle resolved photoemission spectroscopy (ARPES) to study
the momentum dependence of the superconducting gap in
NdFeAsO$_{0.9}$F$_{0.1}$ single crystals. We find that the $\Gamma$
hole pocket is fully gapped below the superconducting transition
temperature. The value of the superconducting gap is 15 $\pm$ 1.5 meV
and its anisotropy around the hole pocket is smaller than 20$\%$ of
this value - consistent with an isotropic or anisotropic s-wave symmetry of the order parameter. This is a significant departure from the situation in
the cuprates, pointing to the possibility that the superconductivity in the iron
arsenic based system arises from a different mechanism.
\end{abstract}


\maketitle The gap function is the single most important quantity that can be used to reveal the
pairing mechanism of a superconductor. It's symmetry and shape in momentum
space are intimately linked to the many body interactions that are responsible
for the creation of the Cooper pairs. The recent
discovery of superconductivity in iron arsenic based materials \cite{Kamihara_original,Takahashi43K,Ren_55K,Rotter2} has initiated intense
experimental \cite{Chen,la_Cruz_NeutronScattering_SDW, Hunte,Jia_PE,Liu_PE, FENG,CHANGLIU1,CHANGLIU2,
Hashimoto_FullGap,YangBaK_ARPES,LiuSrK_ARPES} and theoretical \cite{Mazin,Kuroki,Yao,Lee,Xi,Fa} effort.
The undoped, non-superconducting systems of both oxygen containing
RFeAsO (R=La, Nd, Sm) and oxygen free AFe$_2$As$_2$ (A=Ba, Sr, Ca), display structural \cite{Kamihara_original,Rotter1,NiNiBa,Yan,NiNiCa} and magnetic \cite{la_Cruz_NeutronScattering_SDW,WANG_SDW} phase transitions at elevated temperatures. Doping with RFeAsO with fluorine (electron doping) or AFe$_2$As$_2$ with potassium (hole doping) leads to a suppression of the transition temperature and the emergence of superconductivity \cite{Kamihara_original,Takahashi43K,Ren_55K,Rotter2,NiNiBa}. Perhaps most remarkably, it has recently been discovered \cite{Milton} that undoped CaFe$_2$As$_2$ can also be made superconducting by applying a very
modest amount of external pressure $\sim5$ kbar. One of the most pressing
questions is whether the mechanism of the superconductivity in this
system is similar to that in the classical low temperature
superconductors or the cuprate high temperature superconductors, or if this is a completely new route to the superconducting state. The large atomic
masses of iron and arsenic, combined with the very high critical
temperature, seem to exclude conventional phonon mediated pairing. 
It should be pointed out again, though the superconducting phase transition appears to be in close proximity to a suppressed structural phase transition.  This immediately brings the possibility of phonons and their role in Cooper pair formation back to the forefront.  The fact that in pure CaFe2As2 this can be seen as function of modest pressures, raises the possible role of phonons even further, even though the lack of light elements in the common FeAs-layers requires a subtly enhanced electron-phonon coupling between Fe-$3d$ electrons and vibrations of out of plane atoms. 
A knowledge of the symmetry and shape in momentum space of the superconducting order parameter is essential for constructing the correct model of the pairing mechanism. A number of different scenarios have been proposed to explain the mechanism of the superconductivity in this system \cite{Mazin,Kuroki,Yao,Lee,Xi,Fa}, with predictions about the
symmetry of the order parameter ranging from isotropic and
anisotropic $s$-wave to $d$-wave and $p$-wave. Angle resolved
photoemission spectroscopy (ARPES) is an excellent tool to address
this question. Even though this technique is not sensitive to the
phase of the order parameter, it can directly measure its absolute value as a 
function of momentum via the superconducting gap. In most cases, one
can deduce the character of the order parameter from the symmetry of
the superconducting gap.

Here we report on ARPES measurements of the superconducting gap in
NdFeAsO$_{0.9}$F$_{0.1}$ single crystals. First and most importantly we
find that the Fermi surface is fully gapped - that is, there are
\textit{no} nodes in the order parameter at the Fermi
momenta. The magnitude of the superconducting gap is 15 $\pm$ 1.5 meV
and is comparable to that from our previous measurements within the
experimental uncertainty (error bars and doping) \cite{CHANGLIU1}.
Our data also limits the possible anisotropy of the superconducting
gap to, at most 20\%. Although the results have sizable error bars,
if the gap is indeed anisotropic, the data are consistent with minima located along $\Gamma$ - M, that is at a 45$^\circ$ angle to the Fe-Fe bond. Our results are also consistent with an isotropic $s$-wave symmetry of the order parameter and exclude ordinary $p$-wave and $d$-wave symmetries. However, an anisotropic s-wave state, where gap nodes are located between distinct Fermi surface sheets is also consistent with our data, in particular as such a state creates a small anisotropy of the superconducting gap

High pressure synthesis of samples with the nominal composition of
NdFeAsO$_{0.9}$F$_{0.1}$ was carried out in a cubic, multianvil
press, with an edge length of 19 mm from Rockland Research
Corporation. Stoichiometric amounts of NdFe$_{3}$As$_{3}$,
Nd$_{2}$O$_{3}$, NdF$_{3}$ and Nd were pressed into a pellet with a
mass of approximately 0.5 g and placed inside of a BN crucible with
an inner diameter of 5.5 mm. The synthesis was carried out at a
pressure of 3.3 GPa.  The temperature was increased, over one hour,
from room temperature to 1350 - 1400$^\circ$C and then held there for
8 hours before being quenched to room temperature.  The pressure was
then released and the sample was removed mechanically.  This
synthesis produced a high density pellet that contained large grains
(up to 300 $\times$ 200 $\mu$m in cross section \cite{Ruslan2}) of superconducting ($T{_\textrm{c}}$ $\sim$ 53 K) NdFeAsO$_{0.9}$F$_{0.1}$ as well as non-superconducting NdOFeAs. In addition there are inclusions of
FeAs and Nd$_{2}$O$_{3}$. Magneto optical measurements \cite{Ruslan1}
indicate that on average the samples are over 50\% superconducting.
The single crystals were extracted mechanically from the pellet.
Samples with a size of $\sim200\times200\times50\mu$m were cleaved
\textit{in situ} yielding a flat mirror-like surface. ARPES
experiments were carried out using a Scienta SES2002 hemispherical
analyzer attached to the PGM beam line at the Synchrotron Radiation
Center (SRC), Wisconsin. The profile of the photon beam on the sample
surface was slightly elliptical with a mean diameter smaller than
$\sim100\mu$m. All spectra were measured at 20K using 22 eV photon.
As a reference for the Fermi energy, we used the spectral edge position
of evaporated Au in electrical contact with the sample. The
momentum resolution was set at $0.13^\circ$ and the energy
resolution was set at $\sim16$meV - confirmed by measuring energy
width between 90$\%$ and 10$\%$ intensity positions of Au Fermi
edge.

\begin{figure}
\includegraphics[width=3.6in]{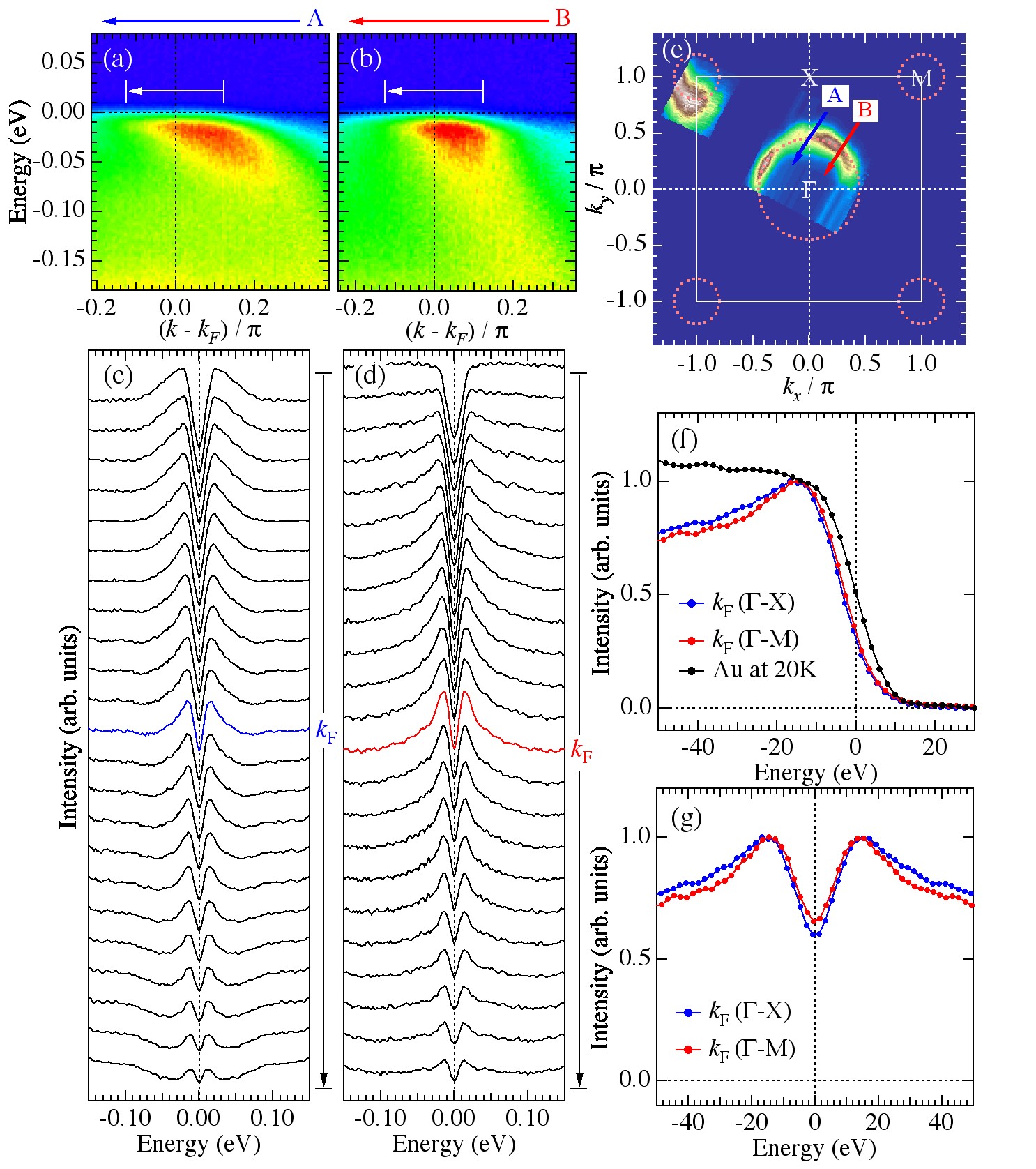}
\caption{{(Color online) The superconducting gap along the high symmetry
directions measured at $T = 20$K. 
(a)-(b) ARPES intensity map along the $\Gamma$-X and $\Gamma$-M directions, respectively  
(directions shown in panel (e)). 
(c)-(d) Energy distribution
curves (EDCs) for panels (a) and (b), respectively. The momentum range is
indicated by the white arrows in panels (a) and (b). The
colored curves mark the EDC at the Fermi momenta. (e)
ARPES intensity map as a function of $k_x$ and $k_y$ momentum,
integrated within 20 meV of the Fermi energy.  Bright areas mark
the location of the Fermi surface. (f) Comparison of the gap magnitude
along $\Gamma$-X and $\Gamma$-M. 
(g) Symmetrized EDCs from the data in panel (f).}} \label{fig1}
\end{figure}

\begin{figure}
\includegraphics[width=3.6in]{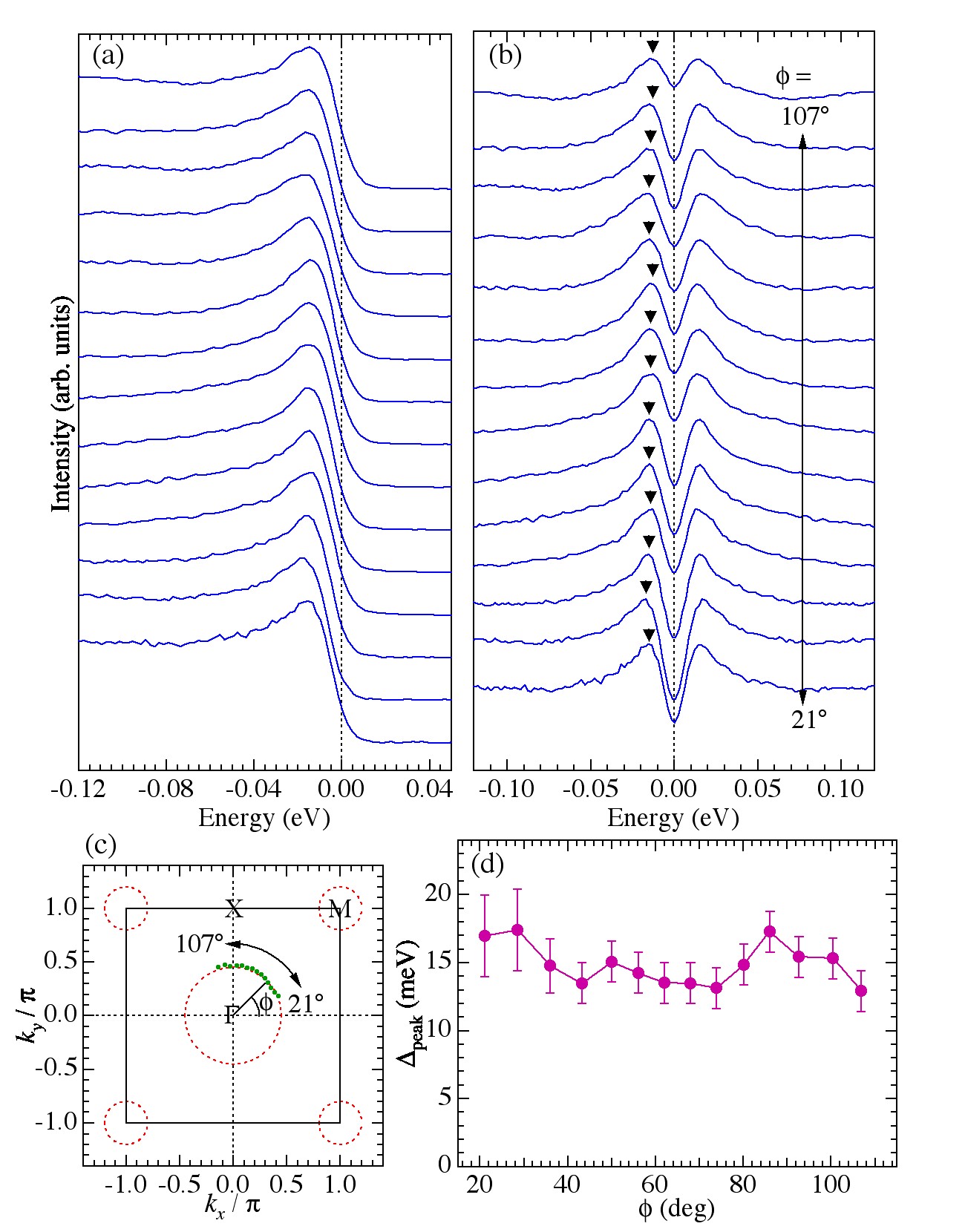}
\caption{(Color online) The magnitude of the superconducting gap along
the $\Gamma$ hole-pocket. (a) raw- and (b) symmetrized-EDCs at the Fermi
crossing momenta marked by green dots in panel (c), where the
definition of the $\phi$ angle is shown. (d) The value of the superconducting gap
extracted from the data in panel (b) using the coherent peak position method.}
\label{fig2}
\end{figure}

We have determined the orientation of the sample and the location of
the Fermi surface from both the ARPES intensity and momentum
distribution curves (MDC).  A plot of the ARPES intensity integrated
over $\pm$ 20 meV about the Fermi energy as a function of momentum \cite{HelenFS,JoelFS} is shown in Fig. 1(e). It reveals a Fermi
surface consistent with our previous report \cite{CHANGLIU1} for
samples from a different batch. Now we focus on two different
momentum cuts (A and B) along the high symmetry directions, $\Gamma$-X
and $\Gamma$-M, respectively. Their location in the Brillouin zone is indicated
by arrows in Fig.1(e). The ARPES intensity along these two cuts is
shown in Figure 1a and b as a function of momentum and energy for $T
= 20$K, deep in the superconducting state. At first glance these
plots exhibit all the characteristics of a sample in the
superconducting state: a buildup of intensity just below the chemical
potential with a characteristic arc shape that arises from particle-hole
mixing \cite{CampuzanoPH} - a hallmark of a pairing induced energy gap.
To confirm this observation of a superconducting gap we used the symmetrization
method \cite{Norman_nature} on the raw EDC data: EDCs are reflected about
the Fermi energy and added to the unreflected ones. This removes the
effects of the Fermi function and enables us to immediately identify
the presence of an energy gap by the appearance of two peaks separated
by a dip, as opposed to a single peak at the chemical
potential when there is no superconducting gap. Figure 1 (c) and (d)
show the symmetrized EDCs corresponding to the data in panel (a) and
(b), respectively. We determined the Fermi wavevector ($k_\textrm{F}$)
from the peak position of the MDCs at the Fermi energy. The opening
of a superconducting gap is clearly observed in both directions: two
sharp peaks do not merge into a single peak but remain separated at
and beyond $k_\textrm{F}$ due to particle hole
mixing \cite{CampuzanoPH}. The width of the coherent peak
($\sim20$meV) is mostly limited by the experimental energy
resolution. Note that the behavior of the coherent peak contrasts
with the rapid broadening of the spectral line shape towards the higher binding
energy due to an increase in the electron scattering
rate \cite{Haule_PRL}. In order to compare the gap sizes along the
two different directions, we plot the EDCs in Fig.1 (f) and the corresponding
symmetrized EDCs in Fig.1 (g). The superconducting gap can be easily
estimated from the EDC data by evaluating the energy position of the
coherent peak \textit{vs}. the Fermi energy. We found that the
superconducting gap has a similar value $\sim$17 $\pm$ 1.5 meV in
both directions. The gap along $\Gamma$-X appears slightly bigger than along the $\Gamma$-M direction (all data points of the peak are further from the chemical potential and the dip is larger), however we cannot conclude this with great certainty given the experimental error bars.

\begin{figure}
\includegraphics[width=3.5in]{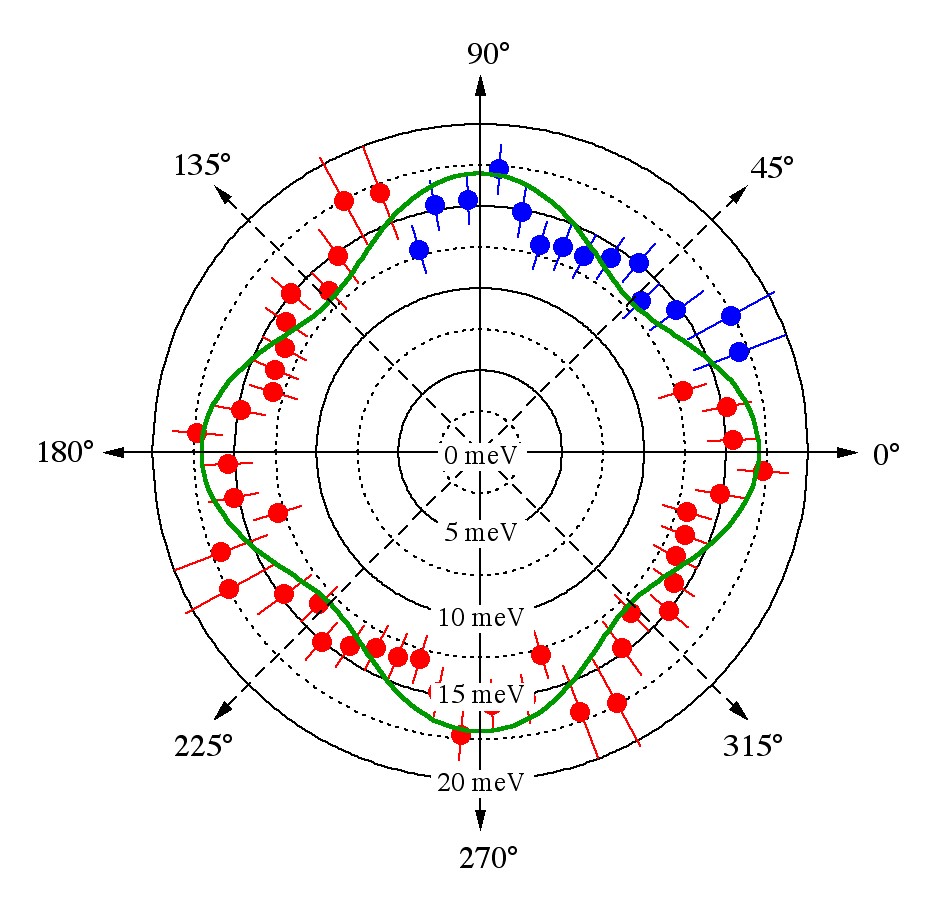}
\caption{(Color online) Magnitude of the superconducting gap around the
$\Gamma$-pocket in polar coordinates. It is clear from this graph that
the gap is always open, with a magnitude ($\Delta_{\textrm{peak}}$)
the varies between $\sim$13meV and $\sim$18meV, indicating conventional
$s$-wave or slightly anisotropic $s$-wave behavior
at the $\Gamma$-pocket. The green line indicates a model with a slight gap
anisotropy of 20$\%$ that would still be consistent with this data.
} \label{fig3}
\end{figure}

We measured the ARPES spectra at a several Fermi momenta points
around the $\Gamma$-centered hole pocket ($\Gamma$-pocket), in
order to obtain the symmetry of the superconducting gap (Fig. 2). We covered
a wide range of the Fermi surface angle ($21^\circ  \le \phi  \le
107^\circ$), which is roughly a quarter of the Fermi surface as shown
in Fig. 2(c). The EDCs and the corresponding symmetrized ones
measured at each Fermi momentum point are shown in Fig. 2(a) and
(b), respectively. We determined the size of the energy gap using
the previously mentioned coherent peak position method. The results are
shown in Fig. 2(d) with the summary shown in polar coordinates in
Fig 3. To give a better sense of the gap symmetry we reflected the
results from one quadrant into the other three quadrants using
the crystal symmetry axes. We use blue symbols to mark the measured data
and red ones to indicate the data points that are a reflection of the
actual data. The superconducting gap is never zero around the
Fermi surface, indicating a lack of nodes. This excludes simple $p$-
or $d$-wave pairing scenarios, which have nodes on the Fermi
surface. In the simplest scenario, our data is consistent with isotropic
$s$-wave behavior, however we cannot exclude the possibility of a
small anisotropy being present of order of 20$\%$ due to the finite
error bars.This would also be consistent with a  pairing
state with nodes of the gap-function between distinct sheets of the Fermi
surface. Such an anisotropic gap is indicated in Fig. 3 by the
green line which shows a 20$\%$ anisotropy and lies within
the error bars of our experiment \cite{Ruslan2}. 
Microscopically, a pairing state with nodes between the Fermi surface sheets is most likely based upon a non-phononic mechanism with strong interband
scattering. A likely candidate mechanism is based upon paramagnon
fluctuations with an in plane wave vector close to $\left( \pi /a,\pi
/a\right)$ \cite{Mazin}, i.e. with a wave vector equal or close to
that of the ordered spin density wave state of the undoped
systems \cite{Qiu}. On the other hand, a fully isotropic s-wave state
would make the electron-phonon mechanism a viable candidate. To distinguish
between the subtle signatures of this limited subset of models, an
additional ARPES study with significantly smaller error bars is necessary.
However, our data already clearly excludes pairing states with gap nodes on
the Fermi surface.

In conclusion, we used angle resolved photoemission spectroscopy to
study the momentum dependence of the superconducting gap in the newly discovered
electron-doped oxypnictide superconductor NdFeAsO$_{0.9}$F$_{0.1}$.
We found a nodeless superconducting gap in the hole pocket around
$\Gamma$ (0,0). The gap magnitude ($\sim15$meV) is almost constant
around the Fermi surface within a variation of less than 20$\%$ ($\sim1.5$meV) - if the gap is indeed anisotropic.
Our results exclude $p$-wave and $d$-wave pairing states with nodes of the gap on the Fermi surface, but are consistent with both an isotropic gap or a state where nodes are located between distinct Fermi surface sheets. The latter, unconventional pairing state implies a small anisotropy of the superconducting gap, consistent with the data presented in this letter. 
Our results are in general agreement with the conclusions of penetration depth measurements \cite{Malone, Ruslan2}.

We acknowledge useful discussions with R. Prozorov and M. A. Tanatar. We thank Helen Fretwell for useful remarks and corrections. Work at Ames Laboratory was supported by the Department of Energy - Basic
Energy Sciences under Contract No. DE-AC02-07CH11358. The
Synchrotron Radiation Center is supported by NSF DMR 9212658. AFSS
thanks LPEM for financial support.

{\it Note added:} After completion of this work we become aware of a
draft manuscript reporting measurements on the hole doped relative, Ba$_{1-x}$K$_x$Fe$_2$As$_2$ \cite{ZhaoGap_ARPES} which also concluded the presence of node-less gap with a consistent magnitude and limits (error bars) on its possible anisotropy. 

%%%%%%%%%%%%%%%%%%%%%%

\end{document}